\documentclass[12pt]{article}
\usepackage{amsmath}
\usepackage{graphicx,psfrag,epsf}
\usepackage{enumerate}
\usepackage{apacite}
\usepackage{natbib}
\usepackage{url}
\usepackage{amsmath,latexsym,amssymb,amsthm}
\newcommand{\blind}{1}
\usepackage{sectsty}
\usepackage{layouts}
\usepackage{epigraph,xcolor}
\usepackage{graphicx}
\usepackage{multirow}
\usepackage{hyperref}

\hypersetup{colorlinks=true,citecolor=blue,
	pdfpagemode=FullScreen,linktocpage=true}

\usepackage{setspace}
\begin{document}
	
\date{}

\if 1\blind
{
\title{\bf Improving linear quantile regression for replicated data}
\author{Kaushik Jana$^{1}$ and Debasis Sengupta$^{2}$\\ $^1$ Imperial College London, UK\\
		$^2$Indian Statistical Institute, Kolkata, India}
	\maketitle
} \fi

\bigskip

\begin{abstract}
\textcolor{black}{This paper deals with improvement of linear quantile regression, when there are a few distinct values of the covariates but many replicates. On can improve asymptotic efficiency of the estimated regression coefficients by using suitable weights in quantile regression, or simply by using weighted least squares regression on the conditional sample quantiles. The asymptotic variances of the unweighted and weighted estimators coincide only in some restrictive special cases, e.g., when the density of the conditional response has identical values at the quantile of interest over the support of the covariate. The dominance of the weighted estimators is demonstrated in a simulation study, and through the analysis of a data set on tropical cyclones.}
\end{abstract}

\section{Introduction}
Consider a quantile regression problem with a handful of distinct values of covariates, where each covariate profile is replicated many times.  Data of this nature arise in various application areas. For instance, intensity of tropical cyclones in a particular season have been studied in relation to the year of occurrence and other year-specific climate variables \citep{Nature_2008, Jagger_2009, Kossin_2013}. Daily/subdaily extreme precipitation levels have been related to monthly El Nino Southern Oscillation (ENSO) index values \citep{Nicholls1993, Ropelewski2008}. \textcolor{black}{In growth research, it is common for scientists to study the dependence of certain quantiles of stature or other body dimensions on the age, ethnicity, gender etc. of the subject, by using cross-sectional data collected on birthdays of subjects \citep{Redden_2004, Fernandez_2004}.} There has also been analysis of stock returns above a certain threshold in a given time period (say in a year), using the time period as covariate \citep{Embrechts1997}.

Given the paucity of distinct values in the covariate space, a linear regression model for the quantiles is often preferred for such data. If one ignores the fact of replications, the linear quantile regression estimator of \cite{Koenker_1978} can be used for estimating the parameters and related inference.  Many of the publications mentioned above have done that.  \textcolor{black}{When the objective is to fit regression models simultaneously for several quantiles, one can use other methods such as Hogg's estimator \citep[p.168]{Koenker2005} or composite quantile regression \citep{zou2008}, which rely on a weighted combination of objective functions for different quantiles.} However, replications provide an opportunity to do better. For instance, the density of the response (at any chosen quantile) can be estimated for each value of the covariate, to get an idea of the likelihoods at different values of the covariates. One can use a weighted version of the linear quantile regression estimator, with larger weight given to more likely values \citep{Knight}.

The replicated nature of the data also enables one to compute a conditional sample quantile for each set of values of the covariates, and fit a linear (mean) regression model to them. Since these conditional sample quantiles would in general have different variances, a weighted least squares estimator with weights inversely proportional to the estimated variances of the respective conditional sample quantiles may be used.

We show in this paper that \textcolor{black}{the two weighted estimators mentioned above are asymptotically more efficient than the estimator of \cite{Koenker_1978}, and characterize conditions for no gain in efficiency. Small sample simulations are conducted to chart the domain of this dominance relation, and an illustrative data analysis is carried out to demonstrate the gains made.}

\section{Comparison of asymptotic variances}
\textcolor{black}{Consider independent sets of data of the form (${\bf x}_i,Y_{ij}$) with $j=1,\ldots,n_i$, $i=1,\ldots,k$. It is assumed that for a given covariate profile ${\bf x}_i$, the observations $Y_{i1},\ldots,Y_{in_i}$ are conditionally iid with common distribution $F_i$. 
Suppose the $\tau$-quantile of $F_i$ is $q(\tau|{\bf x}_i):= \inf\{y : P(Y_{i1} \leq y |{\bf x}_i) \geq \tau \}$. For a fixed $\tau\in [0,1]$, we assume the linear regression model \citep{Koenker2005}
\begin{equation}\label{model1}
q (\tau|{\bf x}_i) = {\bf x}_i' {\boldsymbol\beta}(\tau),
\end{equation}
where ${\boldsymbol\beta}(\tau)$ is the vector of regression coefficients corresponding to the vector of regressors ${\bf x}_i$ (including intercept).}

The sample $\tau$-quantile of the observations for given ${\bf x}_i$ is
\begin{equation}\label{model2}
	\widehat  q(\tau|{\bf x}_i)=\mathop{\arg\min}_{m_i} \sum_{j=1}^{n_i}\rho_{\tau}(Y_{ij}-m_i),\;\;\;i=1,\ldots,k,
\end{equation}
where $\rho_{\tau}(u)=u(\tau-I(u<0))$. We assume that the distribution $F_i$ has continuous Lebesgue density, $f_i$, with $f_i(u) > 0$ on $\{u: 0 < F_i(u) < 1\}$, for $i=1,\ldots,k$.  The limiting distribution of $\widehat q(\tau|{\bf x}_i)$ has mean $q (\tau|{\bf x}_i)$ and variance \citep{Shorack_2009}
\begin{equation}\label{weigts}
	\sigma^2_i(\tau)=\frac{\tau(1-\tau)}{n_i f_i^2(F^{-1}_i(\tau))},\;i=1,\ldots,k.
\end{equation}
Linear regression of $\widehat q (\tau|{\bf x}_i)$ on ${\bf x}_i$, with $\sigma_i^{-2}(\tau)$ as weights, {\bf for $i=1,\ldots,k$}, produces the weighted least squares (WLS) estimator of ${\boldsymbol \beta}(\tau)$
\begin{eqnarray}\label{wls}
	\widehat {\boldsymbol \beta}_{wls}(\tau)
	=({\bf X}'\Omega^{-1}_{\tau}{\bf X})^{-1}{\bf X}'\Omega^{-1}_{\tau}\widehat {\bf q}(\tau)
\end{eqnarray}
where ${\bf X}=({\bf x}_1:\ldots:{\bf x}_k)'$,  $\widehat{\bf q}(\tau)=(\widehat q(\tau|{\bf x}_1),\ldots,\widehat q(\tau|{\bf x}_k))'$ and $\Omega_{\tau}$ is a diagonal matrix with $\sigma^2_1(\tau),\ldots,\sigma^2_k(\tau)$ as diagonal elements, which have to be replaced by consistent estimates. \textcolor{black}{Note that the weight of $\widehat q (\tau|{\bf x}_i)$ in the WLS estimator is the reciprocal of its variance, which depends on $\tau$.}

The estimator proposed by \cite{Koenker_1978} is
\begin{equation}\label{kb}
	\widehat {\boldsymbol \beta}_{kb}(\tau)=\underset{\beta\in \mathbb{R}^2}{\arg\min}\sum_{i=1}^k\sum_{j=1}^{n_i}
	\rho_\tau(Y_{ij}-{\bf x}'_i\boldsymbol{\beta}(\tau)).
\end{equation}
This estimator (the KB estimator) works even if $n_i=1$ for some or all $i$.

In order to show that \eqref{wls} is asymptotically more efficient than \eqref{kb}, we need the following regularity conditions.\\
{\bf Condition A1.}
For some vector $(\xi_{1},\xi_{2},\dots,\xi_{k})^T$ with positive components,
\begin{equation}\label{condE}
	 \left(\frac{n_1}{n},\frac{n_2}{n},\ldots,\frac{n_k}{n}\right)^T\rightarrow \left(\xi_{1},\xi_{2},\dots,\xi_{k}\right)^T
\end{equation}
in Euclidean norm, as $n=\sum_{i=1}^kn_i\rightarrow\infty$.\\
{\bf Condition A2.}
The distribution functions $F_i$ are absolutely continuous, with continuous density $f_i$ uniformly bounded away from $0$ and $\infty$ at  $F^{-1}_i(\tau)$.\\
{\bf Condition A3.}
$\underset{i=1,\ldots,k}{\max}||{\bf x}_i||/\sqrt{n}\rightarrow 0$ as $n\rightarrow\infty$. Further, the sample matrices
$D_{0n}=n^{-1}\!\sum_{i=1}^k{n_i} {\bf x}_i {\bf x}^{T}_i$,  $D_{1n}=n^{-1}\!\sum_{i=1}^k{n_i} f_i(F^{-1}_i(\tau)) {\bf x}_i {\bf x}^{T}_i$ and $D_{2n}=n^{-1}\!\sum_{i=1}^k{n_i} f^2_i(F^{-1}_i(\tau)) {\bf x}_i {\bf x}^{T}_i$ converge to positive definite matrices $D_0$, $D_1$ and $D_2$, respectively, as $n\rightarrow\infty$.

{\bf Theorem 1:}
{\it Under Conditions A1, A2 and A3, and assuming the $\Omega_{\tau}$ in \eqref{wls} is replaced by a consistent estimator,
	\begin{itemize}
		\item [(a)] $\sqrt{n}(\widehat{\boldsymbol \beta}_{kb}(\tau)-{\boldsymbol \beta}(\tau))\rightarrow \mathcal{N}\left(0,\tau(1-\tau)D^{-1}_1 D_0 D^{-1}_1\right),$
		\item [(b)] $\sqrt{n}(\widehat{\boldsymbol \beta}_{wls}(\tau)-{\boldsymbol \beta}(\tau))\rightarrow \mathcal{N}\left(0,\tau(1-\tau) D^{-1}_2\right),$
		\item [(c)] the limiting dispersion matrix of $\sqrt{n}(\widehat{\boldsymbol \beta}_{kb}(\tau)-{\boldsymbol \beta}(\tau))$
		is larger than or equal to that of $\sqrt{n}(\widehat{\boldsymbol \beta}_{wls}(\tau)-{\boldsymbol \beta}(\tau))$ in the sense of the L$\ddot{o}$wner order
		\footnote{A symmetric matrix $A$ is said to be greater than or equal to another symmetric matrix $B$ in the sense of the L$\ddot{o}$wner order if $A-B$ is a non-negative definite matrix.}.
	\end{itemize}}
	\vspace{.1cm}
	{\bf Proof:}
	The result of part (a) follows from \citet[p.121]{Koenker2005}. Part (b) follows from the fact that the WLS estimator is a linear function of the conditional sample quantiles $\widehat q(\tau|{\bf x}_i)$, $i=1,\ldots,k$, whose limiting distribution under the given conditions are well known \citep{Shorack_2009}. The continuous mapping theorem ensures that a consistent estimator of $\Omega_{\tau}$ would be an adequate substitute for it.
	
	Note that the asymptotic dispersion matrices of $\sqrt{n}(\widehat{\boldsymbol \beta}_{kb}(\tau)-{\boldsymbol \beta}(\tau))$ and
	$\sqrt{n}(\widehat{\boldsymbol \beta}_{wls}(\tau)-{\boldsymbol \beta}(\tau))$ are the limits of $\tau(1-\tau)D^{-1}_{1n} D_{0n} D^{-1}_{1n}$ and $\tau(1-\tau)D^{-1}_{2n}$, respectively, where $D_{0n}$, $D_{1n}$ and $D_{2n}$ are as defined in Condition A3.
	Thus, part (c) is proved if we can show that for every $n$, $D^{-1}_{2n}\le D^{-1}_{1n} D_{0n} D^{-1}_{1n}$ in the sense of the L\"{o}wner order. It suffices to show that $D_{1n} D^{-1}_{0n} D_{1n} \le D_{2n}.$
	
	Let $D_{0n}=n^{-1}B'B$, $D_{1n}=n^{-1}A'B=n^{-1}B'A$ and $D_{2n}=n^{-1}A'A$, where
	\begin{equation}\label{matrixab}
		B=\begin{bmatrix} \sqrt{n_1} & \cdots &0\\ \vdots & \ddots&\vdots\\0 &\cdots& \sqrt{n_k}\end{bmatrix}{\bf X},\ \ A=\begin{bmatrix} \sqrt{n_1}f_1(F^{-1}_1(\tau)) & \cdots &0\\
			\vdots & \ddots&\vdots\\0 &\cdots& \sqrt{n_k}f_k(F^{-1}_k(\tau))\end{bmatrix}{\bf X}.
	\end{equation}
	It follows that $$D_{1n} D^{-1}_{0n} D_{1n} = n^{-1}A'B(B'B)^{-1}B'A =n^{-1} A'P_BA \le n^{-1} A'A = D_{2n},$$
	where $P_B$ is the orthogonal projection matrix for the column space of $B$. Part~(c) is proved by taking limits of the two sides of the above inequality as $n$ goes to infinity.\hfill$\Box$


	The next theorem provides a necessary and sufficient condition for the L\"{o}wner order of part (c) to hold with equality.
	
	{\bf Theorem 2:}
	{\it Suppose Conditions A1, A2 and A3 hold and assume that $\Omega_{\tau}$ in \eqref{wls} is replaced by a consistent estimator.
		\begin{itemize}
			\item [(a)] The asymptotic dispersion matrices of the estimators \eqref{wls} and \eqref{kb} coincide if all
			$f_i(F^{-1}_i(\tau))$'s in \eqref{weigts} for $i=1,\ldots,k$ are equal.
			\item [(b)] Suppose ${\bf x}_i=(1: {\bf z}_i')'$ for $i=1,\ldots,k$, where ${\bf z}_1,\ldots,{\bf z}_k$ are samples from a $p$-variate continuous
			distribution not restricted to any lower dimensional subspace. The asymptotic dispersion matrices of the estimators \eqref{wls} and \eqref{kb}
			coincide only if all $f_i(F^{-1}_i(\tau))$'s in \eqref{weigts} for $i=1,\ldots,k$ are equal.
		\end{itemize}
	}
	{\bf Proof:}
	For simplicity of notation, we refer to $f_i(F^{-1}_i(\tau))$ simply by $f_i$ in this proof. The point of departure of the proof of this theorem is part (c) of Theorem~1, where a L$\ddot{\mbox{o}}$wner order between the two dispersion matrices has been established. This order follows from the inequality at the end of the proof of that theorem, which holds with equality if and only if the column space of $A$ is contained in the column space of $B$. From the definition of $A$ and $B$ given in \eqref{matrixab}, this condition amounts to the containment of the column space of ${\bf FX}$ in that of ${\bf X}$, where ${\bf F}$ is the diagonal matrix with $f_1,\ldots,f_k$ as its diagonal elements.
	
	Part (a) is proved by using the fact that if all the $f_i$'s are equal, then ${\bf FX}$ is a constant multiple of ${\bf X}$, implying the equivalence of the column spaces of these two matrices.
	
	In order to prove part (b), we start from the assumption that the column space of ${\bf FX}$ is contained in that of ${\bf X}$, that is, there is a $(p+1)\times(p+1)$ matrix ${\bf C}$ such that ${\bf XC}'=\bf{FX}$. By writing this matrix equation in terms of equality of the corresponding rows of the two sides, we have
	$${\bf C}{\bf x}_i=f_i{\bf x}_i\quad \mbox{for $i=1,\ldots,k$}.$$ Therefore, every $f_i$ is an eigen value of the $(p+1)\times (p+1)$ matrix ${\bf C}$ with eigen vector ${\bf x}_i$. Lemma~1 proved below implies that all the $f_i$'s have to be the same almost surely over the distribution of the ${\bf z}_i$'s mentioned in the statement of the theorem.\hfill$\Box$
	
	{\bf Lemma 1:}
	{\it Suppose ${\bf z}_1,\ldots,{\bf z}_k$ are samples from a p-variate continuous distribution
		not restricted to any lower dimensional subspace. If ${\bf C}$ is a $(p+1)\times(p+1)$ matrix with
		$(1: {\bf z}_1)',\ldots,(1: {\bf z}_{k})'$ as eigen vectors,
		then ${\bf C}$ is almost surely a multiple of the $(p+1)\times(p+1)$ identity matrix.}
	
	{\bf Proof:} Suppose ${\bf z}_1,\ldots,{\bf z}_{p+1}$ are samples drawn initially as in the statement of the lemma and ${\bf C}$ is a $(p+1)\times(p+1)$ matrix
	having $(1: {\bf z}_1)',\ldots,(1: {\bf z}_{p+1})'$
	as eigen vectors. If ${\bf C}$ is not a multiple of the identity matrix, no eigen value of ${\bf C}$ has multiplicity
	$(p+1)$. Therefore, the eigen space (space of eigen vectors) corresponding to each eigen value has dimension $p$ or less.
	For $(1: {\bf z}_{p+2})',\ldots,(1: {\bf z}_k)'$ to be eigen vectors of ${\bf C}$, they have to
	belong to the union of these eigen spaces (each with dimension $<p$). This event has probability zero, according to the hypothesis
	of the lemma. The result follows.\hfill$\Box$
	
	\bigskip
	{\bf Remark 1:}
	The condition $f_1(F^{-1}_1(\tau))=\cdots=f_k(F^{-1}_k(\tau))$ mentioned in Theorem~2 may occur when, for instance,
	the model \eqref{model1} arises from the more restrictive observation model
	\begin{equation}
	Y_{ij}=\beta_0+\beta_1X_i+e_{ij},\; j=1,\ldots,n_i,\; i=1,\ldots,k,    
	\label{eq:restrictive}
	\end{equation}
	where $e_{ij}\sim F$ for some common distribution $F$ that does not depend on $X_i$. This is a special case of \eqref{model1} with
	$\beta_0(\tau)=\beta_0+F^{-1}(\tau)$ and $\beta_1(\tau)=\beta_1$ for all $\tau$.
	By denoting $\mu_i=\beta_0+\beta_1 X_i$, we get $F_i(y)=F(y-\mu_i)$ and $f_i(y)=f(y-\mu_i)$.
	Thus, the conditional $\tau$- quantile is $F_i^{-1}(\tau)=F^{-1}(\tau)+\mu_i$ and the value of the conditional density at that
	quantile is $f_i(F_i^{-1}(\tau))=f(F_i^{-1}(\tau)-\mu_i)=f(F^{-1}(\tau))$,
	for $i=1,\ldots,k$. The equality holds for all $\tau$, which is a much stronger condition than the conditions of Theorem~2.
	
	\bigskip
	In order to define the estimator \eqref{wls} completely, one has to choose a consistent estimator of $\Omega_{\tau}$, which may be obtained by plugging any consistent estimator of $1/(f_i(F_i^{-1}(\tau))$ in \eqref{weigts}. Let us denote $s_i(\tau)=1/(f_i(F_i^{-1}(\tau))$ and
	consider some consistent estimators of this parameter under various conditions.
	
	A simple plug-in estimator is obtained by using the sample quantile to estimate $F_i^{-1}$ and the kernel density estimator
	\citep{Silverman_1986} of $f_i$, for each $i$. If $h_{n_i}$ is the kernel bandwidth, then this estimator would be consistent as long as
	$h_{n_i}\rightarrow0$ and $n_ih_{n_i}\rightarrow\infty$ as $n_i\rightarrow\infty$, and the conditions of Theorem 1 hold.
	
	By noting that $s_i(\tau)=\frac{d}{dt}F_i^{-1}(\tau)$, \cite{siddiqui} proposed the finite difference estimator
	\begin{equation}\label{s_t}
		\hat{s}_i(\tau)=[\widehat q(\tau+h_{n_i}|{\bf x}_i)-\widehat q(\tau-h_{n_i}|{\bf x}_i)]/{2h_{n_i}},
	\end{equation}
	which has been quite popular. This estimator is consistent under the conditions of Theorem~1 when
	the bandwidth parameter $h_{n_i}$ tends to 0 as $n_i\rightarrow\infty$.
	A bandwidth rule, suggested by \cite{Hall_88} for the purpose of obtaining confidence intervals of the $\tau$-quantile based on
	Edgeworth expansions is
	\begin{equation}
		h_{n_i}=n_i^{-1/3} z_{\alpha}^{2/3}[1.5 s_i(\tau)/s_i^{''}(\tau)]^{1/3},\nonumber
	\end{equation}
	where $z_\alpha$ satisfies $\Phi(z_\alpha)=1-\frac{\alpha}{2}$, and $1-\alpha$ is the specified coverage probability of the said confidence
	interval. In the absence of any information about $s_i(\cdot)$, one can use the Gaussian model, as in \cite{Koenker_1999}, to choose
	\begin{equation}\label{bandwidth}
		h_{n_i}=n_i^{-1/3} z_{\alpha}^{2/3}[1.5 \phi^2(\Phi^{-1}(\tau))/(2(\Phi^{-1}(\tau))^2+1)]^{1/3}.
	\end{equation}

\bigskip
{\bf Remark 2:}
A weighted version of Koenker-Basset (WKB) estimate is \begin{equation}\label{wkb}
\widehat {\boldsymbol \beta}_{wkb}(\tau)=\underset{\beta\in \mathbb{R}^2}{\arg\min}\sum_{i=1}^k\sum_{j=1}^{n_i}\widehat{w}_i(\tau)\rho_\tau(Y_{ij}-{\bf x}'_i\boldsymbol{\beta}(\tau)),
\end{equation}
where $\widehat{w}_i(\tau)\propto \widehat {f}_i(q(\tau|{\bf x}_i))$. This estimator gives more weight to those covariate values for which the conditional likelihood is higher at the chosen quantile. \cite{Knight} has shown in an unpublished work that under the assumption of independent and identically distributed response variable, the WKB estimator is first order equivalent to the WLS estimator (and is neither uniformly better nor uniformly worse than it in second order). Thus, the first order dominance of the KB estimator by the WLS estimator also extends to dominance by the WKB estimator.

\section{Simulations of performance}
 We now compare the small sample performances of the estimators $\widehat {\boldsymbol \beta}_{wls}(\tau)$, $\widehat {\boldsymbol \beta}_{kb} (\tau)$, and $\widehat {\boldsymbol \beta}_{wkb}(\tau)$ defined in \eqref{wls}, \eqref{kb}, and \eqref{wkb} in terms of their empirical Mean Squared Error (MSE). \textcolor{black}{We use the WLS estimator defined by \eqref{wls} with $\Omega_{\tau}$ replaced by $$\widehat{\Omega}_{\tau}=\begin{pmatrix} \frac{1}{n_1}\tau(1-\tau)\hat{s}_1(\tau)&0&\cdots&0\\	 0&\frac{1}{n_2}\tau(1-\tau)\hat{s}_2(\tau)&\cdots&0\\ \vdots&\vdots&\ddots&\vdots\\0&0&\cdots&\frac{1}{n_k}\tau(1-\tau)\hat{s}_k(\tau)\\ \end{pmatrix},$$ where $\hat{s}_i(\tau)$ is defined as in \eqref{s_t} together with \eqref{bandwidth} and $\alpha=0.05$. This coincides with the choice made by several other researchers \cite[p.77]{Koenker2005} for estimating the variance of the KB estimator.}

We simulate $k$ distinct values of a scalar covariate in two ways: (a)~as a sample of size $k$ from the gamma distribution with shape parameter $p=2$ and scale parameter $\theta=0.5$ and (b)~as the natural numbers $1,2,\ldots,k$. The later case resembles the cyclone wind speed data analysed in Section~\ref{data-analysis}. Then, for every index $i$ in the range $1,2,\ldots,k$, every index $j$ in the range $1,\ldots,n_i$ and every covariate value $x_i$ chosen as above, we simulate $Y_{ij}$ from \textcolor{black}{the normal distribution} $\mathcal{N}(\mu_i,\eta^2_i)$ where $\mu_i=\beta_0+\beta_1 x_i-\eta_i\Phi^{-1}(\tau)$, so that the $\tau$-quantile of $Y_{ij}$ is $\beta_0+\beta_1x_i$. As for $\eta_i^2$, we choose two different values: $\eta_i=1/x_i$ and $\eta_i=1$. Only the second choice ensures asymptotic equivalence of the two estimators as per Theorem~2. We use $\beta_0=1$, $\beta_1=0.5$, quantile $\tau=0.1$, 0.3, 0.5, 0.7 and 0.9 and number of distinct covariate values $k=10$ and 30. As for the number of replicates $n_i$ of the covariate value $x_i$, we choose the balanced design $n_1=\cdots=n_k$, and use the values 30, 50, 100, 500 and 1000 for $n_1$.  These choices of $\tau$, $k$ and $n_i$ mostly cover the data analytic problems of \cite{Redden_2004}, \cite{Fernandez_2004}, \cite{Nature_2008}, \cite{Jagger_2009} and \cite {Kossin_2013}.

	
We compute the KB estimator \eqref{kb} and the  WKB estimator \eqref{wkb} by using the quantile regression package quantreg (R package version 5.29; //www.r-project.org).
	
Tables~\ref{tab1} and \ref{tab3} show the empirical MSE of the WLS, the KB and the WKB estimators of the two regression parameters, for $\eta_i=1/x_i$ and the specified values of the other parameters, based on 10,000 simulation runs. Tables~\ref{tab1} and \ref{tab3} correspond to Gamma distributed and fixed integer covariates, respectively. \textcolor{black}{The Cramer-Rao lower bound (CRLB) of the regression coefficients under the chosen simulation model are the same for all quantiles, and are reported only in the case of $\tau=0.5$ to avoid repetition. The signal-to-noise ratio (SNR; see page 156 of \cite{Ahmed2019}) for each experimental set-up, which does not depend on $n_1$, is reported below the value of $k$.} 

It can be seen that the empirical MSE of the WLS and the WKB estimators are generally less than that of the KB estimator in the cases of both discrete and continuous covariates. The WLS estimator has larger MSE for the extreme quantiles ($\tau=0.1$ or 0.9) and small sample size ($n_1=$30 and 50 and $k=$10 and 30). This may be because $n_1=$30 and 50 are too small for the estimation of variance of extreme quantiles. For all other cases (higher sample size, less extreme quantile or both), the WLS and the WKB estimators have comparable MSE that is smaller than that of the KB estimator. These small sample findings nicely complement the large sample superiority of the WLS and the WKB estimators over the KB estimator, as described in~Theorem~1 and Remark~2. \textcolor{black}{As the sample size increases, MSE of all the estimators reduce. The gap between the MSE of WLS and WKB estimators and the CRLB also diminish, beginning with the middle quantiles.}
		
We now turn to the case $\eta_i=1$ for all $i$, so that the condition of Theorem~2 holds and the three estimators have asymptotically equivalent performance. Tables~\ref{tab2} and \ref{tab4} show the empirical MSE of the three estimators of the regression parameters based on 10,000 simulation runs, for Gamma distributed and integer covariates, respectively, with other parameters having values specified in the two tables. It is found that there is no clear dominance of any one estimator over the others, for any choice of sample size.

\begin{table}
\caption{Empirical MSE of $\widehat{\boldsymbol\beta}_{kb}$, $\widehat{\boldsymbol\beta}_{wls}$ and $\widehat{\boldsymbol\beta}_{wkb}$ \textcolor{black}{along with CRLB} for $\eta_i=1/x_i$, $i=1,\dots,k$, with Gamma distributed covariate and for different values of $\tau$, $n_1$ and $k$ \textcolor{black}{(with corresponding SNR)}.}
\vspace{0.4cm}
\centering
\resizebox{\columnwidth}{!}{
\begin{tabular}{cccc@{\hskip 5pt}cc@{\hskip 5pt}cc@{\hskip 5pt}cc@{\hskip 5pt}cc@{\hskip 5pt}cc@{\hskip 5pt}c}\hline& $k$& & \multicolumn{2}{c}{$n_1$=30}& \multicolumn{2}{c}{$n_1$=50}&\multicolumn{2}{c}{$n_1$=100} & \multicolumn{2}{c}{\centering $n_1$=500}&\multicolumn{2}{c}{\centering  $n_1$=1000}\\\cline{4-13}
			$\tau$&$\mbox{\textcolor{black}{(SNR)}}$&	$\widehat {\boldsymbol \beta}$ & $\beta_0$& $\beta_1$ &  $\beta_0$& $\beta_1$ & $\beta_0$& $\beta_1$ & $\beta_0$& $\beta_1$ & $\beta_0$& $\beta_1$\\\hline
			0.1&\multirow{3}{*}{10}
			&KB& 0.1591& 0.0783& 0.0877& 0.0521 &0.0426& 0.0250&0.0087& 0.0049& 0.0044&0.0025
			\\&&WLS& 0.1708& 0.0799& 0.0912& 0.0445 &0.0380& 0.0203&0.0063& 0.0037& 0.0031&0.0017
			\\&\textcolor{black}{(4.38)}&WKB& 0.1233& 0.0786& 0.0762& 0.0501 &0.0391& 0.0204&0.0063& 0.0036& 0.0031&0.0018
			\vspace{.1cm}\\
			&\multirow{3}{*}{30}
			& KB& 0.0295& 0.0100& 0.0172& 0.0059 &0.0094& 0.0028&0.0017& 0.0006& 0.0009&0.0003
			\\&&WLS& 0.0655& 0.0135& 0.0299& 0.0064 &0.0084& 0.0025&0.0014& 0.0004& 0.0006&0.0002
			\\&\textcolor{black}{(3.50)}&WKB& 0.0269& 0.0102& 0.0167& 0.0060 &0.0089& 0.0024&0.0013& 0.0005& 0.0007&0.0002
			\\\hline
			0.3&\multirow{3}{*}{10}
			& KB& 0.0842& 0.0487& 0.0494& 0.0286& 0.0247& 0.0139&0.0050& 0.0029&0.0025&0.0015
			\\& &WLS& 0.0655& 0.0360& 0.0373& 0.0213& 0.0184& 0.0102&0.0035& 0.0021&0.0017&0.0010
			\\&\textcolor{black}{(1.94)}& WKB& 0.0640& 0.0363& 0.0375& 0.0214& 0.0179& 0.0112&0.0035& 0.0023&0.0017&0.0011 \vspace{.1cm} \\
			
			&\multirow{3}{*}{30}
			& KB& 0.0172& 0.0058& 0.0102& 0.0035& 0.0051& 0.0017& 0.0010& 0.0004&0.0008&0.0003
			\\&&WLS& 0.0142& 0.0041& 0.0079& 0.0024& 0.0036& 0.0011& 0.0007& 0.0002&0.0006&0.0002
			
			\\&\textcolor{black}{(1.37)}& WKB& 0.0123& 0.0039& 0.0072& 0.0024& 0.0035& 0.0011& 0.0007& 0.0002&0.0006&0.0002\\\hline
			0.5&\multirow{4}{*}{}
			& KB& 0.0764& 0.0456& 0.0459& 0.0278&0.0239& 0.0149& 0.0047& 0.0028&0.0023&0.0013
			\\&10&WLS& 0.0531& 0.0296& 0.0320& 0.0190&0.0154& 0.0090& 0.0031& 0.0018&0.0016&0.0009
			
			\\&\textcolor{black}{(0.92)}& WKB& 0.0553& 0.0308& 0.0332& 0.0193&0.0156& 0.0091& 0.0031& 0.0018&0.0017&0.0010
			\\&&\textcolor{black}{CRLB}& 0.0366& 0.0138& 0.0220& 0.0083&0.0110& 0.0042& 0.0022& 0.0008&0.0011&0.0004 \vspace{.1cm} \\
			&\multirow{4}{*}{}
			& KB &0.0159&0.0053&0.0092& 0.0031&0.0045& 0.0015&0.0009& 0.0003&0.0005&0.0002
			\\&30&WLS   &0.0102&0.0032&0.0062& 0.0019& 0.0030& 0.0009&0.0006& 0.0002&0.0003&0.0001
			
			\\&\textcolor{black}{(0.56)}& WKB&0.0104&0.0033&0.0064& 0.0020&0.0030& 0.0009&0.0009& 0.0002&0.0004&0.0001 \\
			&&\textcolor{black}{CRLB}&0.0126&0.0055&0.0076& 0.0033&0.0038& 0.0016&0.0008& 0.0003&0.0004&0.0002 \\ \hline
			
			0.7&\multirow{3}{*}{10}
			
			& KB & 0.0856& 0.0498& 0.0512 & 0.0314 &0.0248& 0.0175& 0.0049& 0.0028& 0.0025&0.0015
			\\&&WLS& 0.0641& 0.0346& 0.0380 & 0.0218 &0.0181& 0.0010& 0.0034& 0.0019& 0.0017&0.0011
			\\&\textcolor{black}{(0.45)}& WKB   & 0.0628& 0.0352& 0.0380 & 0.0321 &0.0180& 0.0098& 0.0034& 0.0020& 0.0018&0.0011 \vspace{.1cm}\\
			
			&\multirow{2}{*}{30}
			& KB & 0.0171& 0.0057& 0.0104& 0.0035& 0.0051& 0.0017 & 0.0010& 0.0003& 0.0005& 0.0002
			\\&&  WLS& 0.0139& 0.0040& 0.0079& 0.0024& 0.0036& 0.0011 & 0.0007& 0.0002& 0.0003& 0.0001
			
			\\&\textcolor{black}{(0.31)}& WKB& 0.0120& 0.0038& 0.0082& 0.0023& 0.0034& 0.0011 & 0.0007& 0.0002& 0.0003& 0.0001 \\\hline
			0.9&\multirow{3}{*}{10}
			&KB& 0.1453& 0.0874& 0.0870& 0.0490 &0.0802& 0.1000& 0.0089& 0.0051& 0.0044& 0.0015
			\\&&WLS& 0.1731& 0.0822& 0.0912& 0.0432 &0.0612& 0.0717& 0.0068& 0.0038& 0.0043& 0.0011
			\\&\textcolor{black}{(0.74)}&WKB& 0.1246& 0.0807& 0.0780& 0.0424 &0.0236& 0.0701& 0.0064& 0.0037& 0.0043& 0.0011\vspace{.1cm}\\
			&\multirow{3}{*}{30}
			& KB& 0.0294&0.0105& 0.0174& 0.0058 &0.0085& 0.0028&0.0083& 0.0007&0.0007& 0.0005
			\\&&WLS& 0.0656&0.0135& 0.0304& 0.0067 &0.0081& 0.0025&0.0063& 0.0005&0.0005& 0.0004
			\\&\textcolor{black}{(0.91)}& WKB& 0.0268&0.0127& 0.0165& 0.0062 &0.0075& 0.0020&0.0063& 0.0004&0.0004& 0.0004\\\hline
		\end{tabular}
	}\label{tab1}
\end{table}

\begin{table}
	\caption{\textcolor{black}{Empirical MSE of $\widehat{\boldsymbol\beta}_{kb}$, $\widehat{\boldsymbol\beta}_{wls}$ and $\widehat{\boldsymbol\beta}_{wkb}$ along with CRLB ($\times 10^{-3}$)} for $\eta_i=1/x_i$, $i=1,\dots,k$, with integer covariate and for different values of $\tau$, $n_1$ and $k$ (with corresponding SNR).}
	\vspace{0.4cm}
	\centering
	\resizebox{\columnwidth}{!}{
		\begin{tabular}{cccc@{\hskip 5pt}cc@{\hskip 5pt}cc@{\hskip 5pt}cc@{\hskip 5pt}cc@{\hskip 5pt}cc@{\hskip 5pt}c}\hline& $k$& & \multicolumn{2}{c}{ $n_1$=30}& \multicolumn{2}{c}{$n_1$=50}&\multicolumn{2}{c}{$n_1$=100} & \multicolumn{2}{c}{\centering $n_1$=500}&\multicolumn{2}{c}{\centering $n_1$=1000}\\\cline{4-13}
			$\tau$&\mbox{(SNR)}&	$\widehat {\boldsymbol \beta}$ & $\beta_0$& $\beta_1$ &  $\beta_0$& $\beta_1$ & $\beta_0$& $\beta_1$ & $\beta_0$& $\beta_1$ & $\beta_0$& $\beta_1$\\\hline 0.1&\multirow{3}{*}{10}
	&  KB& 6.0782& 0.0944& 3.6725& 0.0572& 1.8733& 0.0290& 0.3675& 0.0057& 0.1823& 0.0029
	\\&&WLS& 6.0901& 0.0886& 3.6043& 0.0488& 1.6292& 0.0229& 0.2749& 0.0041& 0.1264& 0.0019
	\\&(118.80)&WKB& 5.1674& 0.0776& 2.9732& 0.0497& 1.7124& 0.0241& 0.2768& 0.0042 &0.1256& 0.0019 \vspace{.1cm}\\
			
	&\multirow{3}{*}{30}
	& KB& 0.2843& 0.0005& 0.1596& 0.0003& 0.0856& 0.0002&0.0162& 3$\times 10^{-5}$&0.0082&1$\times 10^{-5}$
	\\&&WLS & 0.4786& 0.0006& 0.2213& 0.0003& 0.0826& 0.0001&0.0120& 2$\times 10^{-5}$&0.0054&9$\times 10^{-6}$
	\\&(1805.09)&WKB& 0.2157& 0.0004& 0.1924& 0.0002& 0.0756& 0.0001&0.0107& 2$\times 10^{-5}$&0.0051&9$\times 10^{-6}$ \\\hline
			0.3&\multirow{3}{*}{10}
			& KB& 3.6361& 0.0558& 2.1623& 0.0338& 1.0820& 0.0169& 0.2191& 0.0034& 0.1063& 0.0017
			\\& &WLS& 2.7028& 0.0394& 1.5836& 0.0238& 0.7808& 0.0117& 0.1498& 0.0023& 0.0730& 0.0011
			\\&(109.69)&WKB& 2.7017& 0.0403& 1.5430& 0.0234& 0.7715& 0.0117& 0.1493& 0.0023& 0.0727& 0.0012 \vspace{.1cm} \\
			
			&\multirow{3}{*}{30}
			& KB& 0.1662& 0.0003& 0.0933& 0.0002& 0.0497& $9\times 10^{-5}$&0.0099& $2\times 10^{-5}$& 0.0050&$9\times 10^{-6}$
			\\&&WLS& 0.1214& 0.0002& 0.0663& 0.0001& 0.0348& $6\times 10^{-5}$&0.0061& $1\times 10^{-5}$& 0.0029&$5\times 10^{-6}$
			
			\\&(1785.88)&WKB& 0.1129& 0.0002& 0.0621& 0.0001& 0.0318& $6\times 10^{-5}$&0.0060& $1\times 10^{-5}$&
			0.0029&$8\times 10^{-6}$\\\hline
			0.5&\multirow{4}{*}{}
			&KB& 3.2820& 0.0508& 1.8901& 0.0295& 0.9874& 0.0151& 0.1934& 0.0030&0.0995&0.0016
			\\&10&WLS& 2.3091& 0.0347& 1.2991& 0.0197& 0.6957& 0.0104& 0.1334& 0.0020&0.0670& 0.0010
			\\&(104.05)&WKB& 2.3611& 0.0356& 1.3371& 0.0205& 0.7031& 0.0106& 0.1333& 0.0020&0.0672&0.0010 \\&&CRLB& 1.4014& 0.0213& 0.8408& 0.0128& 0.4204& 0.0064& 0.0841& 0.0013&0.0420&0.0006 \vspace{.1cm} \\
			
			&\multirow{4}{*}{}
			& KB& 0.1511& 0.0003& 0.0877& 0.0002& 0.0447& 9$\times 10^{-5}$&0.0088& 2$\times 10^{-5}$&0.0042&9$\times 10^{-6}$
			\\&30&WLS& 0.0929& 0.0002& 0.0543& 0.0001& 0.0283& 5$\times 10^{-5}$&0.0053& 9$\times 10^{-6}$&0.0027&5$\times 10^{-6}$
			\\&(1773.25)&WKB& 0.0975& 0.0002& 0.0558& 0.0001& 0.0284& 5$\times 10^{-5}$&0.0053& 9$\times 10^{-6}$&0.0027&5$\times 10^{-6}$ 
			\\&&CRLB& 0.0565& 0.0001& 0.0339& 0.0001& 0.0169& 3$\times 10^{-5}$&0.0034& 6$\times 10^{-6}$&0.0017&3$\times 10^{-6}$ \\\hline
			
			0.7&\multirow{3}{*}{10}
			
			&KB& 3.5451&0.0554& 2.1721& 0.0314 &1.0964& 0.0170& 0.2161& 0.0034& 0.1071& 0.0017
			\\&&WLS& 2.6408&0.0391& 1.5121& 0.0221 &0.8168& 0.0123& 0.1485& 0.0022 & 0.0745& 0.0011
			\\&(98.96)&WKB&2.6231&0.0395& 1.1523& 0.0314 &0.8034& 0.0124& 0.1479& 0.0022& 0.0748& 0.0011 \vspace{.1cm}\\
			
			&\multirow{3}{*}{30}
			& KB& 0.1601&0.0003&0.0978& 0.0002&0.0489& 9$\times 10^{-5}$&0.0096& 2$\times 10^{-5}$&0.0049&9$\times 10^{-6}$
			\\&&WLS& 0.1226&0.0002&0.0666& 0.0001&0.0322& 5$\times 10^{-5}$&0.0059& 1$\times 10^{-5}$&0.0029&5$\times 10^{-6}$
			
			\\&(1761.17)&WKB& 0.1187&0.0002&0.0623& 0.0001&0.0303& 5$\times 10^{-5}$&0.0058& 1$\times 10^{-5}$&0.0029&5$\times 10^{-6}$ \\\hline
			0.9&\multirow{3}{*}{10}
			&KB& 6.1749& 0.0957& 3.6214& 0.0565 &1.8597& 0.0287& 0.3593& 0.0055& 0.1836&0.0028
			\\&&WLS& 6.7441& 0.0874& 3.5912& 0.0483 &1.6065& 0.0221& 0.2663& 0.0039& 0.1342&0.0021
			\\&(92.58)&WKB& 5.9431& 0.0798& 2.8872& 0.0436 &1.4269& 0.0208& 0.2581& 0.0039& 0.1341&0.0020
			\vspace{.1cm}\\
				&\multirow{3}{*}{30}
			& KB& 0.2761& 0.0005& 0.1600& 0.0003 &0.0817& 0.0002& 0.0157& 3$\times 10^{-5}$&0.0084&2$\times 10^{-5}$
			\\&&WLS& 0.4901& 0.0006& 0.2231& 0.0003 &0.0810& 0.0001& 0.0131& 2$\times 10^{-5}$&0.0056&1$\times 10^{-5}$
			\\&(1744.70)&WKB& 0.2632& 0.0005& 0.1254& 0.0003 &0.0802& 0.0001& 0.0109& 2$\times 10^{-5}$&0.0052&1$\times 10^{-5}$\\\hline
		\end{tabular}
	}\label{tab3}
\end{table}

\begin{table}
	\caption{Empirical MSE of  $\widehat{\boldsymbol\beta}_{kb}$, $\widehat{\boldsymbol\beta}_{wls}$ and $\widehat{\boldsymbol\beta}_{wkb}$ along with CRLB for $\eta_i=1$, $\forall i$, with Gamma distributed covariate and for different values of $\tau$, $n_1$ and $k$ (with corresponding SNR).}
	\vspace{0.4cm}
	\centering
	\resizebox{\columnwidth}{!}{
		\begin{tabular}{cccc@{\hskip 5pt}cc@{\hskip 5pt}cc@{\hskip 5pt}cc@{\hskip 5pt}cc@{\hskip 5pt}cc@{\hskip 5pt}c}\hline
			& $k$& & \multicolumn{2}{c}{$n_1$=30}& \multicolumn{2}{c}{$n_1$=50}&\multicolumn{2}{c}{$n_1$=100}
			& \multicolumn{2}{c}{\centering $n_1$=500}&\multicolumn{2}{c}{\centering $n_1$=1000}\\\cline{4-13}
			$\tau$&\mbox{(SNR)}&	$\widehat {\boldsymbol \beta}$ & $\beta_0$& $\beta_1$ &  $\beta_0$& $\beta_1$ & $\beta_0$& $\beta_1$ & $\beta_0$& $\beta_1$ & $\beta_0$& $\beta_1$& \\\hline
			
			0.1&\multirow{3}{*}{10}
			
			& KB& 0.0396& 0.0351 &0.0234& 0.0206 &0.0117& 0.0119& 0.0026& 0.0021&0.0011& 0.0010
			\\&&WLS& 0.0596& 0.0399 &0.0318& 0.0231 &0.0113& 0.0128& 0.0025& 0.0022&0.0012& 0.0010
			\\&(8.00)& WKB& 0.0350& 0.0416 &0.0273& 0.0235 &0.0114& 0.0128& 0.0025& 0.0022&0.0012& 0.0010\vspace{.1cm} \\
			
			&\multirow{3}{*}{30}
			&KB& 0.0108& 0.0079& 0.0059& 0.0064 &0.0016& 0.0013&0.0007& 0.0005&0.0003& 0.0002
			\\&&WLS& 0.0283& 0.0108& 0.0038& 0.0069 &0.0012& 0.0013&0.0006& 0.0005&0.0004& 0.0002
			\\&(7.75)&   WKB& 0.0102& 0.0101& 0.0043& 0.0065 &0.0013& 0.0012&0.0006& 0.0005&0.0003&0.0002 \\\hline
			0.3&\multirow{3}{*}{10}
			
			& KB& 0.0236& 0.0209& 0.0142& 0.0125& 0.0070& 0.0062&0.0012& 0.0012& 0.0007 &0.0006
			\\&&WLS& 0.0256& 0.0219& 0.0151& 0.0128& 0.0074& 0.0064&0.0014& 0.0012& 0.0007 &0.0006
			\\&(4.31)&WKB& 0.0254& 0.0225& 0.0151& 0.0132& 0.0074& 0.0064&0.0014& 0.0012& 0.0007 &0.0006 \vspace{.1cm}\\
			&\multirow{3}{*}{30}
			& KB & 0.0065 &0.0048 &0.0039& 0.0027& 0.0018& 0.0013&0.0004& 0.0003&0.0002 &0.0001
			\\&&WLS& 0.0078 &0.0049 &0.0044& 0.0029& 0.0020& 0.0014&0.0004& 0.0003&0.0002 &0.0001
			\\&(4.13)& WKB& 0.0070 &0.0051 &0.0040& 0.0030& 0.0020& 0.0015&0.0004& 0.0003&0.0002 &0.0001\\\hline
			0.5&\multirow{4}{*}{}
			& KB& 0.0213&0.0184&0.0126& 0.0114& 0.0063& 0.0055& 0.0012& 0.0011&0.0006&0.0006
			\\&10&WLS& 0.0216&0.0184&0.0127& 0.0114& 0.0064& 0.0055& 0.0012& 0.0011&0.0007&0.0006
			
			\\&(2.43)&WKB& 0.0217&0.0191&0.0125& 0.0115& 0.0065& 0.0056& 0.0013& 0.0011&0.0006&0.0006
			\\&&CRLB& 0.0145& 0.0098&0.0087& 0.0059& 0.0043& 0.0030& 0.0009& 0.0006&0.0004& 0.0003\vspace{.1cm} \\
			&\multirow{4}{*}{}
			
			& KB& 0.0059&0.0043&0.0034& 0.0025& 0.0017& 0.0012& 0.0003& 0.0002& 0.0002&0.0001
			\\&30&WLS& 0.0060&0.0044&0.0035& 0.0025& 0.0017& 0.0013&0.0003& 0.0002& 0.0002&0.0001
			\\&(2.29)&WKB& 0.0059&0.0043&0.0034& 0.0025& 0.0017& 0.0012& 0.0003& 0.0002& 0.0002&0.0001 \\&&CRLB& 0.0048&0.0039&0.0029& 0.0023& 0.0014& 0.0012& 0.0003& 0.0002& 0.0001&0.0001 \\\hline
			0.7&\multirow{3}{*}{10}
			
			& KB& 0.0229& 0.0201& 0.0151& 0.0121& 0.0069& 0.0070& 0.0013& 0.0012&0.0007&0.0006
			\\&&WLS& 0.0253& 0.0213& 0.0141& 0.0145& 0.0065& 0.0062& 0.0014& 0.0013&0.0007&0.0006
			\\&(1.10)&WKB& 0.0251& 0.0217& 0.0142& 0.0121& 0.0067& 0.0062& 0.0014& 0.0013&0.0007&0.0006\vspace{.1cm}\\
			
			&\multirow{3}{*}{30}
			
			& KB& 0.0065& 0.0046& 0.0039& 0.0039& 0.0017& 0.0014& 0.0004& 0.0003&0.0002&0.0001
			\\&&WLS& 0.0079& 0.0049& 0.0032& 0.0045& 0.0016& 0.0018& 0.0005& 0.0003&0.0002&0.0001
			\\&(1.00)&WKB& 0.0081& 0.0050& 0.0034& 0.0045& 0.0016& 0.0018& 0.0005& 0.0002&0.0002&0.0001\\\hline
			
			0.9&\multirow{3}{*}{10}
			
			& KB& 0.0389& 0.0333& 0.0231& 0.0293 &0.0112& 0.0098& 0.0021& 0.0022&0.0012&0.0010
			\\&&WLS& 0.0565& 0.0379& 0.0215& 0.0226 &0.0141& 0.0118& 0.0024& 0.0021&0.0012&0.0011
			\\&(0.15)&WKB& 0.0477& 0.0395& 0.0222& 0.0246 &0.0132& 0.0112& 0.0020& 0.0021&0.0012&0.0010\vspace{.1cm} \\
			
			&\multirow{3}{*}{30}
			
			& KB& 0.0109& 0.0079& 0.0159& 0.0045 &0.0018& 0.0011& 0.0008& 0.0005&0.0003&0.0002
			\\&&WLS& 0.0285& 0.0108& 0.0136& 0.0053 &0.0021& 0.0014& 0.0007& 0.0005&0.0004&0.0002
			\\&(0.12)&WKB& 0.0247& 0.0098& 0.0151& 0.0050 &0.0017& 0.0013& 0.0007& 0.0005&0.0004&0.0002\\\hline
		\end{tabular}
	}\label{tab2}
\end{table}

\begin{table}
	\caption{Empirical MSE of $\widehat{\boldsymbol\beta}_{kb}$, $\widehat{\boldsymbol\beta}_{wls}$ and $\widehat{\boldsymbol\beta}_{wkb}$ along with CRLB for $\eta_i=1$, $\forall i$, with integer covariate and for different values of $\tau$, $n_1$ and $k$ (with corresponding SNR).}
	\vspace{0.4cm}
	\centering
	\resizebox{\columnwidth}{!}{
		\begin{tabular}{cccc@{\hskip 5pt}cc@{\hskip 5pt}cc@{\hskip 5pt}cc@{\hskip 5pt}cc@{\hskip 5pt}cc@{\hskip 5pt}c}\hline
			& $k$& & \multicolumn{2}{c}{$n_1$=30}& \multicolumn{2}{c}{$n_1$=50}&\multicolumn{2}{c}{$n_1$=100}
			& \multicolumn{2}{c}{\centering $n_1$=500}&\multicolumn{2}{c}{\centering $n_1$=1000}\\\cline{4-13}
			$\tau$&\mbox{(SNR)}&	$\widehat {\boldsymbol \beta}$ & $\beta_0$& $\beta_1$ &  $\beta_0$& $\beta_1$ & $\beta_0$& $\beta_1$ & $\beta_0$& $\beta_1$ & $\beta_0$& $\beta_1$& \\\hline
			
			0.1&\multirow{3}{*}{10}
			& KB& 0.0458& 0.0012 &0.0275& 0.0007 &0.0139& 0.0004& 0.0027& 7$\times 10^{-5}$&0.0014& 4$\times 10^{-5}$
			\\&&WLS& 0.0638& 0.0014 &0.0357& 0.0008 &0.0167& 0.0004& 0.0030& 7$\times 10^{-5}$&0.0014& 4$\times 10^{-5}$
			\\&(27.38)& WKB& 0.0556& 0.0014 &0.0320& 0.0008 &0.0155& 0.0004& 0.0030& 7$\times 10^{-5}$&0.0014& 4$\times 10^{-5}$\vspace{.1cm} \\

			&\multirow{3}{*}{30}
			& KB& 0.0141& 4$\times 10^{-5}$& 0.0081& 2$\times 10^{-5}$& 0.0043&1$\times 10^{-5}$& 0.0008& 3$\times 10^{-6}$&0.0004& 1$\times 10^{-6}$
			\\&&WLS& 0.0321& 6$\times 10^{-5}$& 0.0153& 3$\times 10^{-5}$& 0.0066&2$\times 10^{-5}$& 0.0009& 3$\times 10^{-6}$&0.0004& 1$\times 10^{-6}$
			\\&(119.36)& WKB& 0.0188& 6$\times 10^{-5}$& 0.0121& 3$\times 10^{-5}$& 0.0051&2$\times 10^{-5}$& 0.0009& 3$\times 10^{-6}$&0.0004& 1$\times 10^{-6}$\vspace{.1cm}
			\\\hline
			
			0.3&\multirow{3}{*}{10}
			& KB& 0.0297& 0.0007 &0.0162& 0.0004 &0.0082& 0.0002& 0.0016& 4$\times 10^{-5}$&0.0008& 2$\times 10^{-5}$
			\\&&WLS& 0.0294& 0.0007 &0.0172& 0.0004 &0.0086& 0.0002& 0.0017& 4$\times 10^{-5}$&0.0008& 2$\times 10^{-5}$
			\\&(20.33)& WKB& 0.0296& 0.0007 &0.0173& 0.0004 &0.0086& 0.0002& 0.0017& 4$\times 10^{-5}$&0.0008& 2$\times 10^{-5}$\vspace{.1cm} \\
			
			&\multirow{3}{*}{30}
			& KB& 0.0084& 3$\times 10^{-5}$& 0.0047& 2$\times 10^{-5}$& 0.0025&8$\times 10^{-6}$& 0.0008& 3$\times 10^{-6}$&0.0002& 8$\times 10^{-7}$
			\\&&WLS& 0.0098& 3$\times 10^{-5}$& 0.0054& 3$\times 10^{-5}$& 0.0027&8$\times 10^{-6}$& 0.0009& 3$\times 10^{-6}$&0.0002& 8$\times 10^{-7}$
			\\&(104.74)& WKB& 0.0092& 3$\times 10^{-5}$& 0.0051& 3$\times 10^{-5}$& 0.0026&8$\times 10^{-6}$& 0.0009& 3$\times 10^{-6}$&0.0002& 8$\times 10^{-7}$\vspace{.1cm}
			\\\hline
			
			0.5&\multirow{4}{*}{}
			& KB& 0.0244& 0.0006&0.0142& 0.0004&0.0073& 0.0002& 0.0015& 4$\times 10^{-5}$&0.0007& 2$\times 10^{-6}$
			\\&10&WLS& 0.0247& 0.0006&0.0145& 0.0004&0.0074& 0.0002& 0.0015& 4$\times 10^{-5}$&0.0007& 2$\times 10^{-6}$
			\\&(16.13)& WKB& 0.0254& 0.0007&0.0148& 0.0004&0.0075& 0.0002& 0.0015& 4$\times 10^{-5}$&0.0007& 2$\times 10^{-6}$
			\\&&CRLB&0.0156& 0.0004&0.0093& 0.0002&0.0047& 0.0001&0.0009& 2$\times 10^{-5}$& 0.0005& 1$\times 10^{-5}$\vspace{.1cm} \\
			
			&\multirow{4}{*}{}
			
			& KB& 0.0075& 2$\times 10^{-5}$& 0.0044& 1$\times 10^{-5}$& 0.0022&7$\times 10^{-6}$& 0.0008& 3$\times 10^{-6}$&0.0002& 8$\times 10^{-7}$
			\\&30&WLS& 0.0077& 2$\times 10^{-5}$& 0.0045& 1$\times 10^{-5}$& 0.0023&7$\times 10^{-6}$& 0.0009& 3$\times 10^{-6}$&0.0002& 8$\times 10^{-7}$
			\\&(95.29)& WKB& 0.0079& 3$\times 10^{-5}$& 0.0046& 1$\times 10^{-5}$& 0.0023&8$\times 10^{-6}$& 0.0009& 3$\times 10^{-6}$&0.0002& 8$\times 10^{-7}$			
			\\&&CRLB& 0.0047& 1$\times 10^{-5}$& 0.0028& 9$\times 10^{-6}$& 0.0014&4$\times 10^{-6}$& 0.0003& 1$\times 10^{-6}$&0.0001& 4$\times 10^{-7}$\vspace{.1cm}
			\\\hline
			0.7&\multirow{3}{*}{10}
			
			& KB& 0.0293& 0.0007 &0.0159& 0.0004 &0.0083& 0.0002& 0.0019& 4$\times 10^{-5}$&0.0008& 2$\times 10^{-5}$
			\\&&WLS& 0.0291& 0.0007 &0.0164& 0.0004 &0.0087& 0.0002& 0.0019& 4$\times 10^{-5}$&0.0008& 2$\times 10^{-5}$
			\\&(12.47)& WKB& 0.0292& 0.0007 &0.0166& 0.0004 &0.0087& 0.0002& 0.0019& 4$\times 10^{-5}$&0.0008& 2$\times 10^{-5}$\vspace{.1cm} \\
			
			&\multirow{3}{*}{30}
			
			& KB& 0.0086& 3$\times 10^{-5}$& 0.0050& 2$\times 10^{-5}$& 0.0026&8$\times 10^{-6}$& 0.0008& 3$\times 10^{-6}$&0.0002& 8$\times 10^{-7}$
			\\&&WLS& 0.0093& 3$\times 10^{-5}$& 0.0057& 3$\times 10^{-5}$& 0.0029&8$\times 10^{-6}$& 0.0009& 3$\times 10^{-6}$&0.0002& 8$\times 10^{-7}$
			\\&(86.39)& WKB& 0.0090& 3$\times 10^{-5}$& 0.0055& 3$\times 10^{-5}$& 0.0028&8$\times 10^{-6}$& 0.0009& 3$\times 10^{-6}$&0.0002& 8$\times 10^{-7}$\vspace{.1cm}
			\\\hline
			0.9&\multirow{3}{*}{10}
			
			& KB& 0.0553& 0.0014 &0.0281& 0.0007 &0.0145& 0.0004& 0.0028& 7$\times 10^{-5}$&0.0014& 4$\times 10^{-5}$
			\\&&WLS& 0.0642& 0.0017 &0.0356& 0.0008 &0.0171& 0.0004& 0.0030& 7$\times 10^{-5}$&0.0014& 4$\times 10^{-5}$
			\\&(8.16)& WKB& 0.0587& 0.0017 &0.0331& 0.0008 &0.0169& 0.0004& 0.0031& 7$\times 10^{-5}$&0.0014& 4$\times 10^{-5}$\vspace{.1cm} \\

			&\multirow{3}{*}{30}
			
			& KB& 0.0141& 4$\times 10^{-5}$& 0.0081& 2$\times 10^{-5}$& 0.0044&1$\times 10^{-5}$& 0.0008& 3$\times 10^{-6}$&0.0004& 1$\times 10^{-6}$
			\\&&WLS& 0.0321& 6$\times 10^{-5}$& 0.0153& 3$\times 10^{-5}$& 0.0068&2$\times 10^{-5}$& 0.0009& 3$\times 10^{-6}$&0.0004& 1$\times 10^{-6}$
			\\&(74.51)& WKB& 0.0188& 6$\times 10^{-5}$& 0.0121& 3$\times 10^{-5}$& 0.0052&2$\times 10^{-5}$& 0.0009& 3$\times 10^{-6}$&0.0004& 1$\times 10^{-6}$\vspace{.1cm}
			\\\hline
		\end{tabular}
	}\label{tab4}
\end{table}
	
\section{\textcolor{black}{Analysis of tropical cyclone data}}\label{data-analysis}	
We now use the WLS, KB and WKB estimators to fit model \eqref{model1} to the tropical cyclone data considered in \cite{Nature_2008} and available at {http://myweb.fsu.edu/jelsner/temp /extspace/globalTCmax4.txt}. This satellite based data set consists of lifetime maximum wind speed (metre per second) for each of the 2097 cyclones occurred globally over the years 1981 to 2006. \textcolor{black}{If the year (or any other variable which is observed yearly) is used as covariate, it has 26 distinct values with about 80 replications per year. We start with two covariates: year and annually averaged global sea-surface temperature (SST), available from above website. The focus is on the upper quantiles, representing higher wind speed storms that generally cause major damage.}

\textcolor{black}{In Table~\ref{tab-data-1}, we report the KB estimate (used by \cite{Nature_2008}) of the intercept and slope parameters along with the WLS and the WKB estimates for the cyclone data at the 0.85, 0.9, 0.95 and 0.99 quantiles. We also report the asymptotic standard errors of the three estimators. We observe that the WLS and the WKB estimators have smaller standard error in all the cases. The two covariates have a high degree of linear correlation (larger than 0.81) and one or the other regression coefficient turns out be insignificant at level~0.05. The conflicting signs of the regression coefficients obtained by the different methods in the case of $\tau=0.99$ are explained by the collinearity between the covariates.}

\textcolor{black}{In order to avoid collinearity, we follow \cite{Nature_2008} in choosing the year as the sole variable for explaining quantiles of the maximum wind speed and we report the results in Table~\ref{tab-data-2}. The KB estimator continues to have larger asymptotic standard error than the other two estimators. In particular, the KB estimator of the slope parameter for the quantile $\tau=0.95$ is less than two standard errors away from 0. None of the slope estimators produced by the WLS and WKB methods has this weakness. These findings point towards extreme cyclones becoming progressively more fierce over the years.}

\begin{figure}
		\centering
		\includegraphics[height=5.5 in,width=6.4 in]{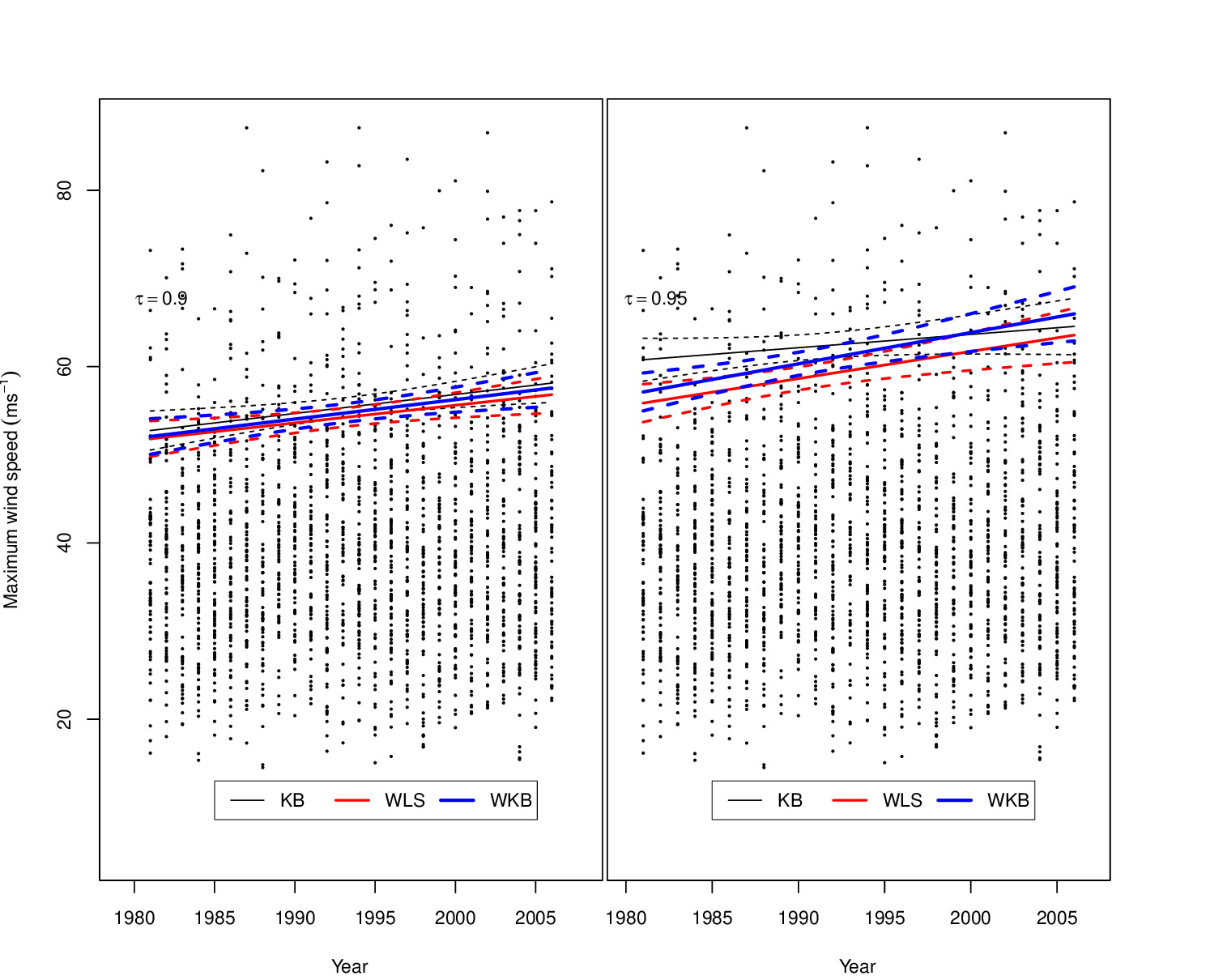}
		\caption{\it Scatter plot of the lifetime maximum wind speeds over the years 1981-2006 along with the regression fit using the KB, WLS and WKB estimators at 0.9 and 0.95 quantiles. The dashed lines represent locus of pointwise confidence limits of the corresponding regression line.}\label{fig1}
	\end{figure}

\textcolor{black}{Figure~\ref{fig1} shows the observed wind speed in successive years and the regression lines along with locus of pointwise confidence limits, obtained from the KB, the WLS and the WKB methods for two chosen quantiles: 0.9 and 0.95. The confidence intervals are narrower for the latter estimators, and they overlap.}

\begin{table}
\caption{KB, WLS and WKB estimators (and their asymptotic standard errors) for different values of $\tau$ for the cyclone lifetime maximum wind speed data with year and SST as covariates.}
\vspace{0.4cm}
\centering

\resizebox{\columnwidth}{!}{
\begin{tabular}{ccccc@{\hskip 20pt}ccc@{\hskip 20pt}ccc@{\hskip 20pt}ccc@{\hskip 20pt}}\hline 
& & \multicolumn{3}{c}{\centering $\tau$=0.85}&\multicolumn{3}{c}{\centering $\tau$=0.90} & \multicolumn{3}{c}{\centering $\tau$=0.95}&  \multicolumn{3}{c}{\centering $\tau$=0.99}\\ \cline{3-14}
&Estimator & $\beta_0$& $\beta_1$& $\beta_2$ & $\beta_0$& $\beta_1$& $\beta_2$ & $\beta_0$& $\beta_1$ & $\beta_2$& $\beta_0$& $\beta_1$& $\beta_2$\\\hline
& KB& -509.85& 0.28213 &-4.1060
&-803.44& 0.43154&-8.14957
&-734.78&0.40094&-10.071
&-819.51&0.44892&-5.28281
\\&&(240.95)&(0.1211)&(3.2195)
&(291.68)&(0.14661)&(3.96312)
&(422.39)&(0.21233)&(6.0423)
&(144.62)&(0.07274)&(2.47813)\\
&WLS& -561.19& 0.3077 &-5.0331& 
-747.73& 0.40288& -6.46277 
&-1162.8&0.61371&-9.93867 
&106.60&-0.01861&10.927
\\&&(232.6801)&(0.1169)&(3.1314)
&(278.66)&(0.14007)&(3.83076)
&(397.16)&(0.19967)&(5.78559)
&(108.58)&(0.05462)&(1.87096)\\
&WKB& -485.87& 0.2700& -3.9934& -802.97&4.30952& -6.94370
& -517.74& 0.2969&-10.644 
&-525.75&0.30098&-0.21887\\
& &(232.68)&(0.1169)&(3.1313)
&(278.66)&(0.14007)&(3.83076)
&(397.16)&(0.19967)&(5.78559)			&(108.58)&(0.05462)&(1.87096)\\ \hline
		\end{tabular}	
	}\label{tab-data-1}
\end{table}

\begin{table}
	\caption{KB, WLS and WKB estimators (and their asymptotic standard errors) for different values of $\tau$ for the cyclone lifetime maximum wind speed data with year as the covariate.}
	\vspace{0.4cm}
	\centering
	\begin{scriptsize}
	\begin{tabular}{cccc@{\hskip 20pt}cc@{\hskip 20pt}cc@{\hskip 20pt}cc@{\hskip 20pt}}\hline 
		& & \multicolumn{2}{c}{\centering $\tau$=0.85}&\multicolumn{2}{c}{\centering $\tau$=0.90} & \multicolumn{2}{c}{\centering $\tau$=0.95}&  \multicolumn{2}{c}{\centering $\tau$=0.99}\\ \cline{3-10}
		
			&Estimator & $\beta_0$& $\beta_1$ & $\beta_0$& $\beta_1$ & $\beta_0$& $\beta_1$ & $\beta_0$& $\beta_1$\\\hline
			
			& KB& -297.66& 0.1754 &-373.56& 0.2152&-238.34& 0.1510&-479.35&0.2778\\&&(140.25)&(0.0703)&(154.93)&(0.0777)&(196.62)&(0.0987)&(67.761)&(0.0340)\\
			
			&WLS& -256.97& 0.1546 &-344.65& 0.2001&-555.70& 0.3087& -438.30 &0.2256 \\&&(135.33)&(0.0679)&(143.40)&(0.0719)& (181.86)&(0.0914)&(55.543)&(0.0279)\\
			
			&WKB& -278.51& 0.1675& -383.96& 0.2201&-644.90& 0.3544& -517.74& 0.2969 \\& &(135.33)&(0.0679)&(143.40)&(0.0719)& (181.86)&(0.0914)&(55.543)&(0.0279)
			\\ \hline
		\end{tabular}	
	\end{scriptsize}
	\label{tab-data-2}
\end{table}

\section{Concluding remarks}

\textcolor{black}{In the case of replicated data, we have shown that the asymptotic efficiency of the usual quantile regression can be improved upon by using suitable weights for replication. Weighted least squares regression of conditional sample quantiles or weighted quantile regression can be useful for this purpose. These improvements are explained by our theoretical results that work for multiple explanatory variables. The asymptotic efficiency of the unweighted estimator coincides with that of the weighted estimators (i.e., using weights is unnecessary) only in very restrictive cases. This can occur, for instance, when the density of the conditional response has identical values at the quantile of interest over the support of the covariate.}

\textcolor{black}{The key to better performance of the WLS or the WKB estimator is their utilization of replications through weights. These weights involve estimation of the density of the response at the given quantile through existing methods \citep{siddiqui, Silverman_1986}. Our simulation study indicates that the cost of estimation of density is not substantial.}

\textcolor{black}{The simulations and a real data analysis conducted here generally support the wisdom of using the WLS or the WKB estimator as an alternative to the KB estimator in the case of replicated data.
}

\bibliographystyle{apalike}
\bibliography{draft}
\end{document}